\documentclass[10pt]{amsart}

\usepackage{amsmath, amscd, amssymb}
\numberwithin{equation}{section}
\newcommand{\sM}{{\mathcal M}}

\newtheorem{theorem}{Theorem}[section]
\newtheorem{Theorem}{Theorem}

\newtheorem{lemma}[theorem]{Lemma}

\theoremstyle{remark}

\theoremstyle{definition}

\begin{document}

\title[Symmetrized cut-join equation of Marino-Vafa formula]
{Symmetrized cut-join equation of Marino-Vafa formula}

\author{Lin Chen}
\address{Department of Mathematics\\
University of California, Los Angeles}
\email{chenlin@math.ucla.edu}

\begin{abstract}In this note, we symmetrized the cut-join equation
from the proof of Marino-Vafa formula. One can derive more
recursion formulas of Hodge integrals out of this polynomial
equations. We also give some applications.
\end{abstract}

\maketitle

\setcounter{tocdepth}{5} \setcounter{page}{1}

\section{Introduction}
The Marino-Vafa formula(Liu-Liu-Zhou's theorem, cf \cite{LLZ1})
gives a close formula for certain Hodge integrals with three
$\lambda$ classes. One of its specialization is the famous ESLV
formula\cite{ESLV}. By applying a transcendental changing of
variable, Goulden,Jackson and Vainshtein get a symmetrized
cut-join equation\cite{GJV}, which is a polynomial identities with
Hodge integral numbers with one $\lambda$ class as coefficients.
Comparing the lowest degree terms, Goulen, Jackson and
Vakil\cite{GJVakil} were able to give a short proof of $\lambda_g$
conjecture,which was first proved by Okounkov and
Pandharipande\cite{FP}. On the other hand, by using the result of
\cite{GJVakil}, Chen, Li and Liu \cite{CLL} gave a short proof of
Witten conjecture Kontsevich theorem.

\vskip 0.4cm

In this paper, we study another transcendental changing of
variable formula, apply it to the Marino-Vafa formula itself, and
get a symmetrized cut-join equation, which is again a polynomial
identity, but with Hodge integrals with three $\lambda$ classes as
coefficients. We expect more Hodge integrals can be computed from
our symmetrized cut-join equation. As an example, we illustrate
how to get the Witten conjecture Kontsevich theorem from our newly
derived symmetrized cut-join equation.

\vskip 0.4cm

We study the new transcendental change of variable formula in
section 2, which is essentially some calculus based on Formal
Lagrange Inversion Theorem. In section 3, we symmetrize the
cut-join equation satisfied by the generating series of Hodge
integral studied in \cite{LLZ1}. Applied the change of variable
formula developed in section 2 to the symmetrized cut-join
equation in section 3, we derived a polynomial cut-join equation,
which is the theorem 3 in section 4, and this is the main result
of this paper. We illustrate some application of our result in
section 5.

\section{Preliminary}
We first quote a result from the standard text book on
combinatorics. For a proof and more about this theorem, we refer
the book Enumerative Combinatorics by Richard Stanley\cite{S}.

\begin{Theorem}
Formal Lagrange Inversion Theorem: Let
$F[x]=\sum_{i=1}^{+\infty}a_ix^i \in xK[[x]]$ where $a_1\neq 0$
and $K$ is a field of characteristic 0. Let $k,n\in
\mathbb{Z}$,then
\begin{equation}
n[x^n]F^{-1}(x)^k=k[x^{n-k}](\frac{x}{F(x)})^n=k[x^{-k}]F(x)^{-n}
\end{equation}
Where $F^{-1}(x)$ denote the formal inverse function of $F(x)$ and
$[x^n]F(x)$ is the coefficient of $x^n$ in the formal power series
$F(x)$.
\end{Theorem}

In particular, take $k=1$, we have
\begin{equation}
n[x^n]F^{-1}(x)=[x^{-1}]F(x)^{-n}
\end{equation}

The inverse function of $x(1-x)^\tau$ will play a crucial role in
this paper. Take $F(x)=x(1-x)^\tau$ in the above theorem for some
fixed complex number $\tau$, then

\begin{equation}
\frac{1}{F(x)^n}=\frac{1}{x^n}(1-x)^{-n\tau}=\frac{1}{x^n}\sum_{r=0}^{+\infty}\frac{\prod_{a=0}^{r-1}(n\tau+a)}{r!}x^r
\end{equation}
and
\begin{equation}
[x^{-1}]\frac{1}{F(x)^n}=\frac{\prod_{a=0}^{n-2}(n\tau+a)}{(n-1)!}
\end{equation}
Let $\omega(x)$ denote $F^{-1}(x)$. We will study some basic
properties of this function in this section. The formal Lagrange
inversion theorem gives the unique formal power series solution of
the equation
\begin{equation}
\omega(x)(1-\omega(x))^\tau=x
\end{equation}

\begin{equation}
\omega(x)=F^{-1}(x)=\sum_{n=1}^{+\infty}\frac{\prod_{a=0}^{n-2}(n\tau+a)}{n!}x^n
\end{equation}

One can compute the derivative
\begin{equation}
x\omega(x)'=\frac{\omega(1-\omega)}{1-(1+\tau)\omega}
\end{equation}

Let $y=\frac{1}{1-(1+\tau)\omega}$, we have

\begin{equation}
(1+\tau)x\omega'=(1-\omega)(\frac{1}{1-(1+\tau)\omega}-1)
\end{equation}

and
\begin{equation}
(1+\tau)^2x\omega'=\frac{\tau}{1-(1+\tau)\omega}+1-(1+\tau)(1-\omega)=\tau
y-\tau+(1+\tau)\omega
\end{equation}

so
\begin{equation}
y=1+(\frac{1+\tau}{\tau})\sum_{n=1}^{+\infty}\frac{\prod_{a=0}^{n-1}(n\tau+a)}{n!}x^n
\end{equation}

For a formal power series $f(x)$, if we change the variable to
$\omega$, and then to $y$, we have the following relations:
\begin{equation}
x\frac{df}{dx}=\frac{1-\omega}{1-(1+\tau)\omega}\omega\frac{df}{d\omega}=y(y-1)(\frac{y\tau+1}{\tau+1})\frac{df}{dy}
\end{equation}

\vskip 0.4cm

\section{Symmetrization}
In \cite{LLZ1}, they studied the generating function
$\mathcal{C}=\sum_{g\geqslant 0,n\geqslant
1}\mathcal{C}_n^g\lambda^{2g-2+n}$, where

\begin{align*}
\mathcal{C}_n^g&=\sum_{d\geqslant 1}\sum_{\mu\vdash
d,l(\mu)=n}-\frac{\sqrt{-1}^{d+n}}{|Aut
\mu|}(\tau(1+\tau))^{n-1}\prod_{i=1}^n\frac{\prod_{a=1}^{\mu_i-1}(\mu_i\tau+a)}{(\mu_i-1)!}\int_{\overline{\sM}_{g,n}}\frac{\Gamma_g(\tau)}{\prod_{i=1}^n(1-\mu_i\psi_i)}\cdot \mathbf{p}_\mu\\
&=-\frac{\sqrt{-1}^{n}(\tau(1+\tau))^{n-1}}{n!}\sum_{\mu_1,\mu_2\cdots\mu_n\geqslant
1}^{+\infty}\sqrt{-1}^{|\mu|}\prod_{i=1}^n\frac{\prod_{a=1}^{\mu_i-1}(\mu_i\tau+a)}{(\mu_i-1)!}\\
& \sum_{k=0}^{3g-3}\sum_{b_1+b_2+\cdots+b_n=3g-3+n-k}\int_{\overline{\sM}_{g,n}}\Gamma_g^k(\tau)\prod_{i=1}^n\psi_i^{b_i}\prod_{i=1}^n\mu_i^{b_i}\cdot \mathbf{p}_\mu\\
&=-\frac{(\tau(1+\tau))^{n-1}}{n!}\sum_{k=0}^{3g-3}\sum_{b_1+b_2+\cdots+b_n=3g-3+n-k}<\tau_{b_1}\cdots\tau_{b_n}\Gamma_g^k(\tau)>\prod_{i=1}^n\phi_{b_i}(\overrightarrow{\mathbf{p}})
\end{align*}

Here we denote
$$\phi_i(\overrightarrow{\mathbf{p}})=\sum_{m\geqslant 1}\sqrt{-1}^{m+1}\mathbf{p}_m\frac{\prod_{a=1}^{m-1}(m\tau+a)}{(m-1)!}m^i=\frac{1}{\tau}\sum_{m\geqslant 1}\sqrt{-1}^{m+1}\mathbf{p}_m\frac{\prod_{a=0}^{m-1}(m\tau+a)}{m!}m^i$$
for infinitely many formal variables
$\overrightarrow{\mathbf{p}}=\{\mathbf{p}_1,\mathbf{p}_2,\cdots\}$
and
$$\Gamma_g(\tau)=\Lambda_g^\vee(1)\Lambda_g^\vee(\tau)\Lambda_g^\vee(-\tau-1)$$.

This apparently complicated generating function naturally appeared
when one computes the open Gromov-Witten invariants of local
Calabi-Yau, cf\cite{KLiu}. Motivated by the duality between
topological string theory and Chern-Simon theory, Marino-Vafa
formula gives a closed expression of the above generating function
$\mathcal{C}$ in terms of some combinatorial data associated to
representations of symmetric groups. In their proof of the
Marino-Vafa formula, Liu-Liu-Zhou show that the generating
function $\mathcal{C}$ satisfies a cut-join equation.

\begin{equation}
\frac{\partial\mathcal{C}}{\partial\tau}=\frac{\sqrt{-1}\lambda}{2}\sum_{i,j\geqslant
1}(ij\mathbf{p}_{i+j}\frac{\partial^2\mathcal{C}}{\partial\mathbf{p_i}\partial\mathbf{p_j}}+ij\mathbf{p}_{i+j}\frac{\partial\mathcal{C}}{\partial\mathbf{p}_i}\frac{\partial\mathcal{C}}{\partial\mathbf{p}_j}+(i+j)\mathbf{p}_i\mathbf{p}_j\frac{\partial\mathcal{C}}{\partial\mathbf{p}_{i+j}})
\end{equation}

In this section we will symmetrize this cut-join equation and make
a transcendental change of the variable, so that the resulting
symmetrized cut-join equation become a polynomial one.

Define the symmetrization operator
$$\Xi_n \mathbf{p}_\alpha=(\sqrt{-1})^{-n-|\alpha|}\sum_{\sigma\in
S_n}x_{\sigma(1)}^{\alpha_1}\cdots x_{\sigma(n)}^{\alpha_n}$$ for
$n\geqslant 1$ if $l(\alpha)=n$ with
$\alpha=(\alpha_1,\cdots,\alpha_n)$, and 0 otherwise.

We have
\begin{equation}
\Xi_n
\mathcal{C}_n^g=-\frac{(\tau(1+\tau))^{n-1}}{n!}\sum_{k=0}^{3g-3}\sum_{b_1+b_2+\cdots+b_n=3g-3+n-k}<\tau_{b_1}\cdots\tau_{b_n}\Gamma_g^k(\tau)>\sum_{\sigma\in
S_n}\prod_{i=1}^n\phi_{b_i}(x_{\sigma(i)})
\end{equation}

where
\begin{equation}
\phi_i(x)=\frac{1}{\tau}\sum_{m\geqslant
1}\frac{\prod_{a=0}^{m-1}(m\tau+a)}{m!}m^ix^m=\frac{1}{\tau+1}(x\frac{d}{dx})^i(y-1)
\end{equation}

Let $\mathbf{C}$ denote the change of variable from $x$ to $y$,
one has the following relation

$$\mathbf{C}x\frac{d}{dx}=y(y-1)(\frac{y\tau+1}{\tau+1})\frac{d}{dy}\mathbf{C}$$
Apply to $\phi_i(x)$ for $i\geqslant 0$, we get

\begin{equation}
\mathbf{C}\phi_i(x)=\mathbf{C}\frac{1}{\tau+1}(x\frac{d}{dx})^i(y-1)=[y(y-1)(\frac{y\tau+1}{\tau+1})\frac{d}{dy}]^i(\frac{y-1}{\tau+1})
\end{equation}

Clearly, this is a polynomial in the new variable $y$ of degree
$2i+1$.

Set
$\Xi^{\{a_1,\cdots,a_m\}}\mathbf{p}_\alpha=\Xi\mathbf{p}_\alpha|_{x_i\rightarrow
x_{a_i},i=1,\cdots,m}$

The following three lemmas are from the section 4 of \cite{GJV}.
However, one should be careful that in our case there are some
extra coefficients appear due to our definition of the symmetrized
operator. We ignore the proof, which one can find in \cite{GJV}.

\begin{lemma}
Let $\alpha$ and $\beta$ be partitions with $l(\alpha)=k$ and
$l(\beta)=m$. Then
$$\Xi^{\{1,\cdots,m+k\}}\mathbf{p}_\alpha\mathbf{p}_\beta=\sum_{(\mathcal{A},\mathcal{B})}(\Xi^{\mathcal{A}}\mathbf{p}_\alpha)(\Xi^{\mathcal{B}}\mathbf{p}_\beta)$$
where the sum is over all ordered partitions
$(\mathcal{A},\mathcal{B})$ of $\{1,\cdots,m+k\}$ with
$|\mathcal{A}|=k$ and $|\mathcal{B}|=m$
\end{lemma}

\begin{lemma}
Let $\alpha$ be a partition $l(\alpha)=m$, and let $1\leqslant
l\leqslant m$, then
\begin{align*}
x_l\frac{\partial}{\partial
x_l}\Xi^{\{1,\cdots,m\}}\mathbf{p}_\alpha
&=\sqrt{-1}^{-m-|\alpha|}\sum_{\sigma\in
S_m}\alpha_{\sigma(l)}\prod_{i=1}^mx_i^{\alpha_{\sigma(i)}}\\
&=\sum_{\sigma\in
S_m}\alpha_{\sigma(l)}\prod_{i=1}^m[\sqrt{-1}^{-1}(-\sqrt{-1}x_i)^{\alpha_\sigma(i)}]\\
&=\sum_{i\geqslant
1}\sqrt{-1}^{-(i+1)}(x_l)^i\Xi^{\{1,\cdots,m\}-\{l\}}i\frac{\partial\mathbf{p}_\alpha}{\partial\mathbf{p}_i}
\end{align*}
\end{lemma}

\begin{lemma}
Let $\alpha$ be a partition with $l(\alpha)=m+1$ and $1\leqslant
l\leqslant m$, then

$$x_l\frac{\partial}{\partial x_l}x_{m+1}\frac{\partial}{\partial
x_{m+1}}\Xi^{\{1,\cdots,m+1\}}\mathbf{p}_\alpha|_{x_{m+1}=x_l}=\sum_{i,j\geqslant
1}\sqrt{-1}^{-(i+j+2)}x_l^{i+j}\Xi^{\{1,\cdots,m+1\}-\{l,m+1\}}ij\frac{\partial^2}{\partial\mathbf{p}_i\partial\mathbf{p}_j}\mathbf{p}_\alpha$$
\end{lemma}

Now we apply the operator $\Xi^{\{1,\cdots,m\}}$ to the cut-joint
equation to get a symmetrized one. Notice that
$\Xi^{\{1,\cdots,m\}}$ commutes with taking derivative with
respect to $\tau$, the left hand side gives

$$\Xi^{\{1,\cdots,m\}}\frac{\partial\mathcal{C}}{\partial\tau}=\frac{\partial}{\partial\tau}(\Xi^{\{1,\cdots,m\}}\mathcal{C})=\sum_{g\geqslant
0}\lambda^{2g-2+m}\frac{\partial}{\partial\tau}(\Xi^{\{1,\cdots,m\}}\mathcal{C}_m^g)$$

Next we study the effect of $\Xi^{\{1,\cdots,m\}}$ on the right
hand side. By lemma 3.3,


\begin{align*}
\Xi^{\{1,\cdots,m\}}\sum_{i,j\geqslant
1}\mathbf{p}_{i+j}ij\frac{\partial^2\mathbf{p}_\alpha}{\partial\mathbf{p}_i\partial\mathbf{p}_j}
&=\sum_{l=1}^m\sum_{i,j\geqslant
1}\sqrt{-1}^{-(i+j+1)}x_l^{i+j}\Xi^{\{1,\cdots,m\}-\{l\}}ij\frac{\partial^2\mathbf{p}_\alpha}{\partial\mathbf{p}_i\partial\mathbf{p}_j}\\
&=\sqrt{-1}\sum_{l=1}^mx_l\frac{\partial}{\partial
x_l}x_{m+1}\frac{\partial}{\partial
x_{m+1}}\Xi^{\{1,\cdots,m+1\}}\mathbf{p}_\alpha|_{x_{m+1}=x_l}
\end{align*}

\begin{align*}
\Xi^{\{1,\cdots,m\}}\sum_{i,j\geqslant
1}\mathbf{p}_{i+j}ij\frac{\partial^2\mathcal{C}}{\partial\mathbf{p}_i\partial\mathbf{p}_j}
&=\sum_{g\geqslant 0}\lambda^{2g-2+m+1}\sum_{i,j\geqslant
1}\Xi^{\{1,\cdots,m\}}\mathbf{p}_{i+j}ij\frac{\partial^2\mathcal{C}_{m+1}^g}{\partial\mathbf{p}_i\partial\mathbf{p}_j}\\
&=\sqrt{-1}\sum_{g\geqslant
0}\lambda^{2g-2+m+1}\sum_{l=1}^mx_l\frac{\partial}{\partial
x_l}x_{m+1}\frac{\partial}{\partial
x_{m+1}}\Xi^{\{1,\cdots,m+1\}}\mathcal{C}_{m+1}^g|_{x_{m+1}=x_l}
\end{align*}


Let $l(\alpha)=k$ and $l(\beta)=m-k+1$, and so we have

\begin{align*}
\Xi^{\{1,\cdots,m\}}\sum_{i,j\geqslant 1}
\mathbf{p}_{i+j}ij\frac{\partial\mathbf{p}_\alpha}{\partial\mathbf{p}_i}\frac{\partial\mathbf{p}_\beta}{\partial\mathbf{p}_j}
&=\sum_{l=1}^m\sum_{i,j\geqslant
1}(\sqrt{-1})^{-(1+i+j)}x_l^{i+j}\Xi^{\{1,\cdots,m\}-\{l\}}(i\frac{\partial\mathbf{p}_\alpha}{\partial\mathbf{p}_i})(j\frac{\partial\mathbf{p}_\beta}{\partial\mathbf{p}_j})\\
&=\sqrt{-1}\sum_{l=1}^m\sum_{(\mathcal{A},\mathcal{B})}(\sum_{i\geqslant
1}\sqrt{-1}^{-(i+1)}x_l^i\Xi^{\mathcal{A}}i\frac{\partial\mathbf{p}_\alpha}{\partial\mathbf{p}_i})(\sum_{j\geqslant
1}\sqrt{-1}^{-(j+1)}x_l^j\Xi^{\mathcal{B}}j\frac{\partial\mathbf{p}_\beta}{\partial\mathbf{p}_j})\\
&=\sqrt{-1}\sum_{l=1}^m\sum_{(\mathcal{A},\mathcal{B})}(x_l\frac{\partial}{\partial
x_l}\Xi^{\mathcal{A}\cup\{l\}}\mathbf{p}_\alpha)(x_l\frac{\partial}{\partial
x_l}\Xi^{\mathcal{B}\cup\{l\}}\mathbf{p}_\beta)\\
&=\sqrt{-1}\Theta_{k-1}(x_1\frac{\partial}{\partial
x_1}\Xi^{\{1,\cdots,k\}}\mathbf{p}_\alpha)(x_1\frac{\partial}{\partial
x_1}\Xi^{\{1,k+1,\cdots,m\}}\mathbf{p}_\beta)
\end{align*}

\begin{align*}
&\Xi^{\{1,\cdots,m\}}\sum_{i,j\geqslant 1}
\mathbf{p}_{i+j}ij\frac{\partial\mathcal{C}}{\partial\mathbf{p}_i}\frac{\partial\mathcal{C}}{\partial\mathbf{p}_j}\\
=&\sum_{g_1,g_2\geqslant 0}\sum_{1\leqslant k\leqslant
m}\lambda^{2g_1-2+k}\cdot\lambda^{2g_2-2+(m-k+1)}\cdot\Xi^{\{1,\cdots,m\}}\sum_{i,j\geqslant
1}
\mathbf{p}_{i+j}ij\frac{\partial\mathcal{C}_k^{g_1}}{\partial\mathbf{p}_i}\frac{\partial\mathcal{C}_{m-k+1}^{g_2}}{\partial\mathbf{p}_j}\\
=&\sum_{g_1,g_2\geqslant 0}\sum_{1\leqslant k\leqslant
m}\lambda^{2g_1+2g_2-3+m}\sqrt{-1}\Theta_{k-1}(x_1\frac{\partial}{\partial
x_1}\Xi^{\{1,\cdots,k\}}\mathcal{C}_k^{g_1})
(x_1\frac{\partial}{\partial
x_1}\Xi^{\{1,k+1\cdots,m\}}\mathcal{C}_{m-k+1}^{g_2})
\end{align*}


\begin{align*}
\Xi^{\{1,\cdots,m\}}\sum_{i,j\geqslant
1}(i+j)\mathbf{p}_i\mathbf{p}_j\frac{\partial\mathbf{p}_\alpha}{\partial\mathbf{p}_{i+j}}
&=2\sum_{1\leqslant l<k\leqslant m}\sum_{i,j\geqslant
1}\sqrt{-1}^{-(i+j+2)}x_l^i x_k^j\Xi^{\{1,\cdots,m\}-\{l,k\}}(i+j)\frac{\partial\mathbf{p}_\alpha}{\partial\mathbf{p}_{i+j}}\\
&=2\sum_{1\leqslant l<k\leqslant
m}\sum_{r\geqslant1}\sqrt{-1}^{-(r+2)}\cdot\frac{x_k^r
x_l-x_l^rx_k}{x_k-x_l}\cdot\Xi^{\{1,\cdots,m\}-\{l,k\}}r\frac{\partial\mathbf{p}_\alpha}{\partial\mathbf{p}_r}\\
&=-2\sqrt{-1}\sum_{1\leqslant l\neq k\leqslant
m}\frac{x_k}{x_l-x_k}x_l\frac{\partial}{\partial
x_l}\Xi^{\{1,\cdots,m\}-\{k\}}\mathbf{p}_\alpha\\
&=-2\sqrt{-1}\Theta_1\frac{x_2}{x_1-x_2}\cdot
x_1\frac{\partial}{\partial
x_1}\Xi^{\{1,3,4\cdots,m\}}\mathbf{p}_\alpha
\end{align*}

\begin{align*}
\Xi^{\{1,\cdots,m\}}\sum_{i,j\geqslant
1}(i+j)\mathbf{p}_i\mathbf{p}_j\frac{\partial\mathcal{C}}{\partial\mathbf{p}_{i+j}}
&=\sum_{g\geqslant
0}\lambda^{2g-3+m}\Xi^{\{1,\cdots,m\}}\sum_{i,j\geqslant
1}(i+j)\mathbf{p}_i\mathbf{p}_j\frac{\partial\mathcal{C}_{m-1}^g}{\partial\mathbf{p}_{i+j}}\\
&=\sum_{g\geqslant
0}-2\sqrt{-1}\lambda^{2g-3+m}\cdot\Theta_1\frac{x_2}{x_1-x_2}\cdot
x_1\frac{\partial}{\partial
x_1}\Xi^{\{1,3,4\cdots,m\}}\mathcal{C}_{m-1}^g
\end{align*}


Collecting all these terms, the following theorem is proved.
\begin{Theorem} The symmetrized cut-join equation of Marino-Vafa
formula is
\begin{align*}
&\sum_{g\geqslant 0}\lambda^{2g-2+m}\frac{\partial}{\partial\tau}\Xi^{\{1,\cdots,m\}}\mathcal{C}_m^g\\
=&-\frac{1}{2}\sum_{g\geqslant
0}\lambda^{2g+m}(\sum_{l=1}^mx_l\frac{\partial}{\partial
x_l}x_{m+1}\frac{\partial}{\partial
x_{m+1}}\Xi^{\{1,\cdots,m+1\}}\mathcal{C}_{m+1}^g)|_{x_{m+1}=x_l}\\
&-\frac{1}{2}\sum_{g_1,g_2\geqslant 0}\sum_{1\leqslant k \leqslant
m}\lambda^{2g_1+2g_2+m-2}\Theta_{k-1}(x_1\frac{\partial}{\partial
x_1}\Xi^{\{1,\cdots,k\}}\mathcal{C}_k^{g_1})(x_1\frac{\partial}{\partial
x_1}\Xi^{\{1,k+1,\cdots,m\}}\mathcal{C}_{m-k+1}^{g_2})\\
&+\sum_g\lambda^{2g+m-2}\Theta_1\frac{x_2}{x_1-x_2}\cdot
x_1\frac{\partial}{\partial
x_1}\Xi^{\{1,3,4\cdots,m\}}\mathcal{C}_{m-1}^g
\end{align*}
\end{Theorem}

Comparing the coefficients of $\lambda^{2g-2+m}$, we get

\begin{align*}
\frac{\partial}{\partial\tau}\Xi^{\{1,\cdots,m\}}\mathcal{C}_m^g
=&-\frac{1}{2}\sum_{l=1}^mx_l\frac{\partial}{\partial
x_l}x_{m+1}\frac{\partial}{\partial
x_{m+1}}\Xi^{\{1,\cdots,m+1\}}\mathcal{C}_{m+1}^{g-1})|_{x_{m+1}=x_l}\\
&-\frac{1}{2}\sum_{0\leqslant a \leqslant g}\sum_{1\leqslant k
\leqslant m}\Theta_{k-1}(x_1\frac{\partial}{\partial
x_1}\Xi^{\{1,\cdots,k\}}\mathcal{C}_k^a)(x_1\frac{\partial}{\partial
x_1}\Xi^{\{1,k+1,\cdots,m\}}\mathcal{C}_{m-k+1}^{g-a})\\
&+\Theta_1\frac{x_2}{x_1-x_2}\cdot x_1\frac{\partial}{\partial
x_1}\Xi^{\{1,3,4\cdots,m\}}\mathcal{C}_{m-1}^g
\end{align*}

\vskip 0.4cm

\section{Changing of variable}


Now we want to make a change of the variable for the equation we
obtained in the last section. We first deal with the right hand
side. As in \cite{GJV}, to obtain a polynomial expression in the
variable $y_i$, one has to combined all the unstable terms, which
are logarithm transcendental.

In the second term, combined the unstable terms $a=0$, $k=1$ and
$a=g$, $k=m$
$$\sum_{l=1}^m(x_l\frac{\partial}{\partial
x_l}\Xi^{\{l\}}\mathcal{C}_l^0)(x_l\frac{\partial}{\partial
x_l}\Xi^{\{1,2,\cdots,m\}}\mathcal{C}_m^g)$$

and recall that (3.2)

\begin{equation}
\Xi^{\{l\}}\mathcal{C}_l^0=\Xi^{\{l\}}\sum_{d=1}^{+\infty}-\sqrt{-1}^{d+1}\frac{\prod_{a=1}^{d-1}(d\tau+a)}{(d-1)!}\cdot
d^{-2}\cdot
\mathbf{p}_d=-\sum_{d=1}^{+\infty}\frac{\prod_{a=1}^{d-1}(d\tau+a)}{d!}\cdot
\frac{x_l^d}{d}
\end{equation}

\begin{equation}
(x_l\frac{\partial}{\partial
x_l})^2\Xi^{\{l\}}\mathcal{C}_l^0=-\frac{1}{\tau}\sum_{d=1}^{+\infty}\frac{\prod_{a=0}^{d-1}(d\tau+a)}{d!}x_l^d=-\frac{y_l-1}{\tau+1}=-\frac{\omega_l}{1-(\tau+1)\omega_l}
\end{equation}

since $x\frac{\partial}{\partial
x}=\frac{1-\omega}{1-(\tau+1)\omega}\cdot
\omega\frac{\partial}{\partial\omega}$ (2.11), we find the unique
expression of

\begin{equation}
x_l\frac{\partial}{\partial
x_l}\Xi^{\{l\}}\mathcal{C}_l^0=\ln(1-\omega_l)
\end{equation}

The unstable terms $a=0$, $k=2$ and $a=g$, $k=m-1$ gives

$$\Theta_1(x_1\frac{\partial}{\partial
x_1}\Xi^{\{1,2\}}\mathcal{C}_2^0)(x_1\frac{\partial}{\partial
x_1}\Xi^{\{1,3,\cdots,m\}}\mathcal{C}_{m-1}^g)$$

and

\begin{align*}
\Xi^{\{1,2\}}\mathcal{C}_2^0&=-\tau(\tau+1)\sum_{\mu_1\geqslant
1,\mu_2\geqslant
1}\frac{x_1^{\mu_1}x_2^{\mu_2}}{\mu_1+\mu_2}\prod_{i=1,2}\frac{\prod_{a=1}^{\mu_i-1}(\mu_i\tau+a)}{(\mu_i-1)!}\\
&=-\frac{\tau+1}{\tau}\sum_{\mu_1\geqslant 1,\mu_2\geqslant
1}\frac{x_1^{\mu_1}x_2^{\mu_2}}{\mu_1+\mu_2}\prod_{i=1,2}\frac{\prod_{a=0}^{\mu_i-1}(\mu_i\tau+a)}{\mu_i!}
\end{align*}

\begin{equation}
(x_1\frac{\partial}{\partial x_1}+x_2\frac{\partial}{\partial
x_2})\Xi^{\{1,2\}}\mathcal{C}_2^0=-\tau(\tau+1)\cdot
\frac{\omega_1\omega_2}{[1-(\tau+1)\omega_1][1-(\tau+1)\omega_2]}
\end{equation}

One can verify that
$$\Xi^{\{1,2\}}=-\ln(\frac{\omega_1-\omega_2}{x_1-x_2})-\tau(\ln(1-\omega_1)+\ln(1-\omega_2))$$
is the unique solution, and so

\begin{align*}
x_1\frac{\partial}{\partial x_1}\Xi^{\{1,2\}}
&=-\frac{\omega_1(1-\omega_1)}{(\omega_1-\omega_2)(1-(\tau+1)\omega_1)}+\frac{x_1}{x_1-x_2}+\frac{\tau\omega_1}{1-(\tau+1)\omega_1}\\
&=-\frac{\omega_1}{\omega_1-\omega_2}(1+\frac{\tau\omega_2}{1-(\tau+1)\omega_1})+\frac{x_1}{x_1-x_2}
\end{align*}

Remember we have $\omega(1-\omega)^{\tau}=x$, $\omega$ depends on
the parameter $\tau$. Taking partial derivative to $\tau$, we find

$$\frac{\partial\omega}{\partial\tau}\cdot\frac{1-(\tau+1)\omega}{\omega(1-\omega)}+\ln(1-\omega)=0$$

Move the terms involve $\ln(1-\omega_l)$ to the left the
symmetrized cut-join equation, since fix $\omega$,$\partial
y(\omega,\tau)/\partial\tau=y^2\omega$, we have

\begin{align*}
\mathbf{C}\{\frac{d}{d\tau}\Xi^{\{1,\cdots,m\}}\mathcal{C}_m^g-\sum_{l=1}^m\frac{\partial\omega_l}{\partial\tau}\cdot\frac{\partial}{\partial\omega_l}\Xi^{\{1,\cdots,m\}}\mathcal{C}_m^g\}
&=\frac{d}{d\tau}\mathbf{C}\Xi^{\{1,\cdots,m\}}\mathcal{C}_m^g(y_1(\omega_1,\tau),\cdots,y_m(\omega_m,\tau),\tau)\\
&=(\frac{\partial}{\partial\tau}+\sum_{l=1}^m\omega_ly_l^2\frac{\partial}{\partial
y_l}))\mathbf{C}\Xi^{\{1,\cdots,m\}}\mathcal{C}_m^g(y_1,\cdots,y_m,\tau)
\end{align*}

\begin{Theorem}
After the transcendental changing of variables to $y$, the
symmetrized generating series
$\mathbf{C}\Xi^{\{1,\cdots,m\}}\mathcal{C}_m^g(y_1,\cdots,y_m,\tau)$
is a polynomial of the variables $y_i$'s of total degree
$6g-6+3m$, and satisfy the following cut-joint equation:
\begin{align*}
(\frac{\partial}{\partial\tau}&+\sum_{l=1}^m\frac{y_l(y_l-1)}{\tau+1}\cdot
\frac{\partial}{\partial
y_l})\mathbf{C}\Xi^{\{1,\cdots,m\}}\mathcal{C}_m^g(y_1,\cdots,y_m,\tau)\\
=&-\frac{1}{2}\sum_{l=1}^my_l(y_l-1)(\frac{y_l\tau+1}{\tau+1})\frac{\partial}{\partial
y_l}\cdot
y_{m+1}(y_{m+1}-1)(\frac{y_{m+1}\tau+1}{\tau+1})\frac{\partial}{\partial
y_{m+1}}\mathbf{C}\Xi^{\{1,\cdots,m+1\}}\mathcal{C}_{m+1}^{g-1}|_{y_{m+1}=y_l}\\
&-\frac{1}{2}\sum_{1\leqslant a\leqslant g-1}\sum_{1\leqslant
k\leqslant
m}\Theta_{k-1}(y_1(y_1-1)(\frac{y_1\tau+1}{\tau+1})\frac{\partial}{\partial
y_1}\mathbf{C}\Xi^{\{1,\cdots,k\}}\mathcal{C}_k^a)\\
&\cdot(y_1(y_1-1)(\frac{y_1\tau+1}{\tau+1})\frac{\partial}{\partial
y_1}\mathbf{C}\Xi^{\{1,k+1,\cdots,m\}}\mathcal{C}_{m-k+1}^{g-a})\\
&-\sum_{k=3}^m\Theta_{k-1}(y_1(y_1-1)(\frac{y_1\tau+1}{\tau+1})\frac{\partial}{\partial
y_1}\mathbf{C}\Xi^{\{1,\cdots,k\}}\mathcal{C}_k^0)(y_1(y_1-1)(\frac{y_1\tau+1}{\tau+1})\frac{\partial}{\partial
y_1}\mathbf{C}\Xi^{\{1,k+1,\cdots,m\}}\mathcal{C}_{m-k+1}^{g})\\
&+\Theta_1\frac{y_1^2(y_1-1)(y_2-1)}{y_1-y_2}(\frac{y_1\tau+1}{\tau+1})\frac{\partial}{\partial
y_1}\mathbf{C}\Xi^{\{1,3,\cdots,m\}}\mathcal{C}_{m-1}^{g}
\end{align*}
\end{Theorem}

\section{Applications}
This cut-join equation is a generalization of the symmetrized
cut-join equation of \cite{GJVakil}. In their equation, only Hodge
integrals with at most one $\lambda$ class show up, while in ours
equation, Hodge integrals have up to three $\lambda$ classes. This
is not surprising, since their starting point ESLV formula, as
Liu-Liu-Zhou showed \cite{LLZ1} and \cite{LLZ2}, is the large
$\tau$ limit of Marino-Vafa formula. Thus we expect taking large
$\tau$ limit of our symmetrized cut-join equation, one should be
able to recover the equation of \cite{GJV}.

\vskip 0.4cm

To illustrate the application of our symmetrized cut-join
equation, we make a similar derivation of the Witten conjecture
(Kontevich Theorem) as in \cite{CLL}. We don't regard this as a
new proof.

Both of the two sides are polynomials of $m$ variables
$y_1,\cdots,y_m$, of total degree $6g-5+3m$. We compare this
leading term. Recall
$\Gamma_g(\tau)=\Lambda_g^\vee(1)\Lambda_g^\vee(\tau)\Lambda_g^\vee(-\tau-1)$,
and only its constant $(-)^g[\tau(\tau+1)]^g$ has contribution in
the leading degree term. Denote $\mathbf{F}_d$ the operator
sending a formal power series to its degree $d$ part.

\begin{align*}
&\mathbf{F}_{6g-6+3m}\mathbf{C}\Xi^{\{1,\cdots,m\}}\mathcal{C}_m^g(y_1,\cdots,y_m,\tau)\\
=&(-1)^{g-1}(\frac{\tau^2}{1+\tau})^{2g-2+m}\sum_{b_1+\cdots+b_m=3g-3+m}<\tau_{b_1},\cdots,\tau_{b_m}>\prod_{i=1}^m(2b_i-1)!!y_i^{2b_i+1}
\end{align*}

Where in the above equation, we adopt the following abbreviation
and the genus $g$ is determined by the restriction
$j_1+\cdots+j_n+d=3g-3+n$ if the degree of $\omega$ is $d$.
\begin{equation}
<\tau_{j_{1}}\cdots\tau_{j_{n}}\omega>:=\int_{\overline{\sM}_{g,n}}\psi^{j_{1}}_{1}\cdots\psi^{j_{n}}_{n}\omega.
\end{equation}
For the left hand side, only the derivatives of $y_i$ have
contributions:

\begin{align*}
&\mathbf{F}_{6g-5+3m}(\frac{\partial}{\partial\tau}+\sum_{l=1}^m\frac{y_l(y_l-1)}{\tau+1}\cdot
\frac{\partial}{\partial
y_l})\mathbf{C}\Xi^{\{1,\cdots,m\}}\mathcal{C}_m^g(y_1,\cdots,y_m,\tau)\\
=&\frac{1}{\tau+1}\sum_{l=1}^my_l^2\frac{\partial}{\partial y_l}\mathbf{F}_{6g-6+3m}\mathbf{C}\Xi^{\{1,\cdots,m\}}\mathcal{C}_m^g(y_1,\cdots,y_m,\tau)\\
=&\frac{(-1)^{g-1}}{1+\tau}(\frac{\tau^2}{1+\tau})^{2g-2+m}\sum_{l=1}^m\sum_{b_1+\cdots+b_m=3g-3+m}<\tau_{b_1},\cdots,\tau_{b_m}>\cdot
(2b_l+1)!!y_l^{2b_l+2}\prod_{i=1,i\neq l}^m(2b_i-1)!!y_i^{2b_i+1}
\end{align*}

Now go to the right hand side, after applying the operator the
$\mathbf{F}_{6g-5+3m}$, the first term becomes
\begin{align*}
\frac{1}{2}\cdot\frac{(-1)^{g-1}}{1+\tau}(\frac{\tau^2}{1+\tau})^{2g-2+m}\sum_{l=1}^m&\sum_{b_1+\cdots+b_{m+1}=3g-5+m}<\tau_{b_1},\cdots,\tau_{b_{m+1}}>\\
&\cdot(2b_l+1)!!(2b_{m+1}+1)!!y_l^{2b_l+2b_{m+1}+6}\prod_{i=1,i\neq
l}^m(2b_i-1)!!y_i^{2b_i+1}
\end{align*}

and the second term becomes.
\begin{align*}
&\frac{(-1)^{g-1}}{2(1+\tau)}(\frac{\tau^2}{1+\tau})^{2g-2+m}\\
&\cdot\sum_{1\leqslant a\leqslant g-1}\sum_{1\leqslant k \leqslant m}
\Theta_{k-1}[\sum_{b_1+\cdots+b_k=3a-3+k}<\tau_{b_1},\cdots,\tau_{b_k}>
(2b_1+1)!!y_1^{2b_1+3}\prod_{i=2}^k(2b_i-1)!!y^{2b_i+1}]\\
&\cdot[\sum_{b'_1+b_{k+1}+\cdots+b_m=3(g-a-1)+(m+1-k)}
<\tau_{b'_1},\tau_{b_{k+1}},\cdots,\tau_{b_m}>(2b'_1+1)!!y_1^{2b'_1+3}
\prod_{i=k+1}^m(2b_i-1)!!y^{2b_i+1}]
\end{align*}

The third term basically is the same as the second, except that
the summation range is fixing $a=0$ and $k$ varies from $3$ to
$m$.

\begin{align*}
&\frac{(-1)^{g-1}}{1+\tau}(\frac{\tau^2}{1+\tau})^{2g-2+m}\\
&\cdot\sum_{3\leqslant k \leqslant m}\Theta_{k-1}[\sum_{b_1+\cdots+b_k=k-3}
<\tau_{b_1},\cdots,\tau_{b_k}>(2b_1+1)!!y_1^{2b_1+3}\prod_{i=2}^k(2b_i-1)!!y^{2b_i+1}]\\
&\cdot[\sum_{b'_1+b_{k+1}+\cdots+b_m=3(g-1)+(m+1-k)}<\tau_{b'_1},\tau_{b_{k+1}},\cdots,\tau_{b_m}>
(2b'_1+1)!!y_1^{2b'_1+3}\prod_{i=k+1}^m(2b_i-1)!!y^{2b_i+1}]
\end{align*}

Together with the second term, these gave all the stable cut
contributions, and we combine them in the sequel.

The fourth term is
\begin{align*}
\frac{1}{2}&\frac{(-1)^{g-1}}{1+\tau}(\frac{\tau^2}{1+\tau})^{2g-2+m}\Theta_1\sum_{b_1+b_3+\cdots+b_m=3g-4+m}<\tau_{b_1},\tau_{b_3},\cdots,\tau_{b_m}>\\
&(2b_1+1)!!y_1y_2(\frac{y_1^{2b_1+4}-y_2^{2b_1+4}}{y_1-y_2})\prod_{i=3}^m(2b_i-1)!!y_i^{2b_i+1}
\end{align*}

Collect all these, and comparing the coefficients of
$y_l^{2b_l+2}\prod_{i=1,i\neq l}^my_i^{2b_i+1}$, we get the
Dijkgraaf-Verlinde-Verlinde formula, which is equivalent to Witten
conjecture. See also \cite{CLL} and \cite{KimL} for more detail.

\vskip 0.4cm

For other more interesting applications, one may take other
special values of $\tau$, or consider other degree in terms in
theorem 3. For example, the lowest and the next lowest degree
terms of theorem 3 may give some relations for Hodge integrals
$\int_{\overline{\sM}_{g,n}}\psi^{j_{1}}_{1}\cdots\psi^{j_{n}}_{n}\lambda_g\lambda_{g-1}\lambda_{g-3}$,
which may be interesting.

\end{document}